\documentclass[runningheads]{llncs}
\usepackage[T1]{fontenc}
\usepackage{tikz, pgfplots, algorithmic, algorithm, url, amsmath}
\usepackage{graphicx}

\begin{document}
	\title{Diffusion Models for Medical Anomaly Detection}
	\titlerunning{Diffusion Models for Medical Anomaly Detection}
	%
	\author{Julia Wolleb, 
		Florentin Bieder, 
		Robin Sandk\"uhler,  
		Philippe C. Cattin}  

	\authorrunning{J. Wolleb et al.}
	
	\institute{Department of Biomedical Engineering, University of Basel, Allschwil, Switzerland\\
		\email{julia.wolleb@unibas.ch}}
	\maketitle              
	\begin{abstract}
		In medical applications,  weakly supervised anomaly detection methods are of great interest, as only image-level annotations are required for training. Current anomaly detection methods mainly rely on generative adversarial networks or autoencoder models. Those models are often complicated to train or have difficulties to preserve fine details in the image. We present a novel weakly supervised anomaly detection method based on denoising diffusion implicit models. We combine the deterministic iterative noising and denoising scheme with classifier guidance for
		image-to-image translation between diseased and healthy subjects. Our method generates very detailed anomaly maps without the need for a complex training procedure.
		We evaluate our method on the BRATS2020 dataset for brain tumor detection and the CheXpert dataset for detecting pleural effusions.
	\end{abstract}
	
	\section{Introduction}
	In medical image analysis, pixel-wise annotated ground truth is hard to obtain, often unavailable and contains a bias to the human annotators. Weakly supervised anomaly detection has gained a lot of interest in research as an essential tool to overcome the aforementioned issues. Compared to fully supervised methods, weakly supervised models rely only on image-level labels for training. In this paper, we present a novel pixel-wise anomaly detection approach based on Denoising Diffusion Implicit Models (DDIMs) \cite{ddim}. 
	\begin{figure}[h!]
		\centering
		\includegraphics[width=1\textwidth]{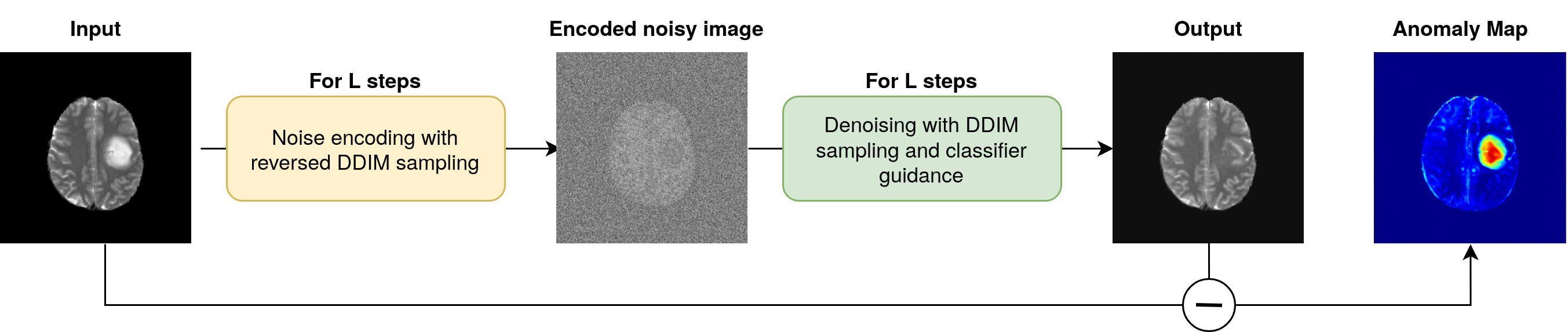}
		\caption{Proposed sampling scheme for image-to-image translation between a diseased input image and a healthy output image. The anomaly map is defined as the difference between the two.}
		\label{overview}
	\end{figure}
	Figure~\ref{overview} shows an overview of the proposed method. We assume two unpaired sets of images for the training, the first containing images of healthy subjects and the second images of subjects affected by a disease. Only the image and the corresponding image-level label (healthy, diseased) are provided during training. \\
	Our method consists of two main parts. In the first part, we train a Denoising Diffusion Probabilistic Models (DDPM)\cite{ddpm} and a binary classifier on a dataset of healthy and diseased subjects. In the second part, we create the actual anomaly map of an unseen image. For this, we first encode the anatomical information of an image with the reversed sampling scheme of DDIMs. This is an iterative noising process. Then, in the denoising process, we use the deterministic sampling scheme proposed in DDIM with classifier guidance to generate an image of a healthy subject. The final pixel-wise anomaly map is the difference between the original and the synthetic image. With this encoding and denoising procedure, our method can preserve many details of the input image that are not affected by the disease while re-painting the diseased part with realistic looking tissue.
	We apply our algorithm on two different medical datasets, i.e., the BRATS2020 brain tumor challenge \cite{brats1,brats2,brats3}, and the CheXpert dataset \cite{chexpert}, and compare our method against standard anomaly detection methods. The source code and implementation details are available at \mbox{\url{https://gitlab.com/cian.unibas.ch/diffusion-anomaly}}.
	
	\subsubsection{Related Work}
	In classical anomaly detection, autoencoders \cite{auto1,kingma} are trained on data of healthy subjects. Any deviations from the learned distribution then lead to a high anomaly score. This idea has been applied for unsupervised anomaly detection in medical images \cite{autobrats1,autobrats2,marimont}, where the difference between the healthy reconstruction and the anomalous input image highlight pixels that are perceived as anomalous.
	Other approaches focus on Generative Adversarial Networks (GANs) \cite{gan} for image-to-image translation \cite{fixed,vagan,descargan}.\\
	However, training of GANs is challenging and requires a lot of hyperparameter tuning. Furthermore, additional loss terms and changes to the architecture are required to ensure cycle-consistent results. In \cite{cam,saliency}, the gradient of a classifier is used to obtain anomaly maps. O
	Recently, transformer networks \cite{transformers2} were also successfully applied on brain anomaly detection \cite{transformers1}. Non-synthesis based methods such as density estimation, feature modeling or self-supervised classification also provide state-of-the-art techniques for anomaly detection \cite{yang2021visual}. In \cite{challenging}, a new thresholding method is proposed for  anomaly segmentation on the BRATS dataset. \\
	Lately, DDPMs were in focus for there ability to beat GANs on image synthesis \cite{beatgans}. In the flow of this success, they were also applied on image-to-image translation \cite{unitddpm,ilvr}, segmentation \cite{diffseg}, reconstruction \cite{palette} and registration\cite{diffusemorph}. As shown in  \cite{ddim}, DDIMs are closely related to score-based generative models \cite{score_based}, which can be used for interpolation between images. However, there is no diffusion model for anomaly detection so far to the best of our knowledge.
	
	\section{Method}
	A typical example for image-to-image translation in medicine is the transformation of an image of a patient to an image without any pathologies.
	For anomaly detection it is crucial that only pathological regions are changed, and the rest of the image is preserved. Then, the difference between the original and the translated image defines the anomaly map. Our detail-preserving image-to-image translation is based on diffusion models. We follow the formulation of DDPMs given in \cite{ddpm,improving}. In Algorithm~\ref{alg:guidingddim}, we present the workflow of our approach.\\
	The general idea of diffusion models is that for an input image $x$, we generate a series of noisy images $\{x_0, x_1, ..., x_T\}$ by adding small amounts of noise for many timesteps $T$. The noise level $t$ of an image $x_t$ is steadily increased from $0$ to $T$. A U-Net $\epsilon_{\theta}$ is trained to predict $x_{t-1}$ from $x_t$ according to \eqref{ddim2}, for any step $t \in \{1,...,T\}$. During training, we know the ground truth for $x_{t-1}$, and the model is trained with an MSE loss. During evaluation, we start from $x_T \sim \mathcal{N}(0,\mathbf{I})$ and predict $x_{t-1}$ for $t \in \{T,...,1\}$. With this iterative denoising process, we can generate a a fake image  $x_{0}$.
	The forward noising process $q$ with variances  $\beta_{1},...,\beta_{T}$ is defined by
	\begin{equation}\label{eq:forward}
	q(x_{t}|x_{t-1}):=\mathcal{N}(x_{t};\sqrt{1-\beta _{t}}x_{t-1},\beta _{t}\mathbf{I}).
	\end{equation}
	This recursion can be written explicitly as 
	\begin{equation}\label{eq:property}
	x_{t}=\sqrt[]{\bar{\alpha} _{t}}x_{0}+\sqrt[]{1-\bar{\alpha} _{t}}\epsilon, \quad \mbox{with } \epsilon \sim \mathcal{N}(0,\mathbf{I}).
	\end{equation}
	with $\alpha _{t}:=1-\beta _{t}$ and $\bar{\alpha}_{t}:=\prod_{s=1}^t \alpha _{s}$. 
	The denoising process $p_{\theta}$ is learned by optimizing the model parameters $\theta$ and is given by 
	\begin{equation}\label{eq:reverse}
	p_{\theta}(x_{t-1}\vert x_t):= \mathcal{N}\bigl(x_{t-1};\mu_{\theta}(x_t, t), \Sigma_\theta(x_t,t)\bigr).
	\end{equation}
	The output of the U-Net is denoted as $\epsilon_{\theta}$, and the MSE loss used for training is
	\begin{equation}\label{eq4}
	\mathcal{L}:= ||\epsilon-\epsilon_{\theta}(\sqrt[]{\bar{\alpha} _{t}}x_{0}+\sqrt[]{1-\bar{\alpha} _{t}}\epsilon, t)||^2_2 , \quad \mbox{with } \epsilon \sim \mathcal{N}(0,\mathbf{I}).
	\end{equation}
	As shown in \cite{ddim}, we use the DDPM formulation to predict $x_{t-1}$ from $x_t$ with
	\begin{equation}\label{ddim2}
	x_{t-1} = \sqrt{\bar{\alpha}_{t-1}}\left(\frac{x_t-\sqrt{1-\bar{\alpha}_t}\epsilon_\theta(x_t,t)}{\sqrt{\bar{\alpha}_t}}\right)+\sqrt{1-\bar{\alpha}_{t-1}-\sigma_t^2}\epsilon_{\theta}(x_t,t)+\sigma_t\epsilon, 
	\end{equation}
	with $\sigma_t = \sqrt{(1 - \bar{\alpha}_{t-1}) / (1 - \bar{\alpha}_t)} \sqrt{1 - \bar{\alpha}_t / \bar{\alpha}_{t-1}}$.  DDPMs have a stochastic element $\epsilon$ in each sampling step \eqref{ddim2}. In DDIMs however, we set ${\sigma_t=0}$, which results in a deterministic sampling process.
	As derived in \cite{ddim}, \eqref{ddim2} can be viewed as the Euler method to solve an ordinary differential equation (ODE). Consequently, we can reverse the generation process by using the reversed ODE. Using enough discretization steps,  we can encode $x_{t+1}$ given $x_t$ with
	\begin{equation}\label{reversed}
	x_{t+1}  =  x_{t}+\sqrt{\bar{\alpha}_{t+1}} \left[ \left( \sqrt{\frac{1}{\bar{\alpha}_{t}}} - \sqrt{\frac{1}{\bar{\alpha}_{t+1}}}\right) x_t + \left(\sqrt{\frac{1}{\bar{\alpha}_{t+1}} - 1} - \sqrt{\frac{1}{\bar{\alpha}_{t} }- 1} \right) \epsilon_\theta(x_t,t) \right].
	\end{equation}
	By applying \eqref{reversed} for $t \in \{0,...,T-1\}$, we can encode an image $x_0$ in a noisy image $x_T$. Then,  we recover the identical $x_0$ from $x_T$ by using \eqref{ddim2} with $\sigma_t=0$ for $t \in \{T,...,1\}$.\\
	For anomaly detection, we train a DDPM on a dataset containing images of healthy and diseased subjects. For evaluation, we define a noise level $L \in \{1, ... , T\}$ and a gradient scale $s$. Given an input image $x$, we encode it to a noisy image $x_{L}$ using \eqref{reversed} for $t \in \{0, ... , L-1\}$. With this iterative noising process, we can induce anatomical information of the input image. During the denoising
	process, we follow \eqref{ddim2} with $\sigma_t=0$ for $t \in \{L, ... , 1\}$.
	We apply classifier guidance as introduced in \cite{beatgans} to lead the image generation to the desired healthy class $h$. For this, we pretrain a classifier network $C$ on the noisy images $x_t$ for $t \in \{1,...,T\}$, to predict the class label of $x$. During the denoising process, the scaled gradient $s  \nabla_{x_t} \log C(h|x_t,t)$ of the classifier is used to update $\epsilon_{\theta}(x_t,t)$. This iterative noising and denoising scheme is  presented in Algorithm \ref{alg:guidingddim}. We generate an image $x_{0}$ of the desired \mbox{class $h$} that preserves the basic structure of $x$. The anomaly map is then defined by the difference between $x$ and $x_{0}$. 
	The choice of the noise level $L$ and the gradient scale $s$ is crucial for the trade-off between detail-preserving image reconstruction and freedom for translation to a healthy subject.
	\begin{algorithm}
		\caption{Anomaly detection using noise encoding and classifier guidance}
		\label{alg:guidingddim}
		\begin{algorithmic}
			\STATE Input: input image $x$, healthy class label $h$, gradient scale $s$, noise level $L$\\
			Output: synthetic image $x_{0}$, anomaly map $a$
			\FORALL{$t$ from 0 to $L-1$ }
			\STATE $x_{t+1} \gets x_{t}+\sqrt{\bar{\alpha}_{t+1}} \left[ \left( \sqrt{\frac{1}{\bar{\alpha}_{t}}} - \sqrt{\frac{1}{\bar{\alpha}_{t+1}}}\right) x_t + \left(\sqrt{\frac{1}{\bar{\alpha}_{t+1}} - 1} - \sqrt{\frac{1}{\bar{\alpha}_{t} }- 1} \right) \epsilon_\theta(x_t,t) \right]$
			\ENDFOR
			\FORALL{$t$ from $L$ to 1}
			\STATE $\hat \epsilon \gets \epsilon_{\theta}(x_t,t) - s\* \sqrt{1-\bar{\alpha}_t}  \nabla_{x_t} \log C(h|x_t,t)$
			\STATE $x_{t-1} \gets \sqrt{\bar{\alpha}_{t-1}} \left( \frac{x_t - \sqrt{1-\bar{\alpha}_t} \hat{\epsilon}}{\sqrt{\bar{\alpha}_t}} \right) + \sqrt{1-\bar{\alpha}_{t-1}} \hat{\epsilon}$
			\ENDFOR
			\STATE $a \gets \sum\limits_{channels} \bigl|x-x_{0} \bigr|$
			\RETURN $x_{0}$, $a$
		\end{algorithmic}
	\end{algorithm}
	
	\section{Experiments}
	The DDPM is trained as proposed in \cite{improving} without data augmentation.
	We choose the hyperparameters for the DDPM model as described in the appendix of \cite{beatgans}, for $T=1000$ sampling steps. The model is trained with the Adam optimizer and the hybrid loss objective described in \cite{improving}, with a learning rate of ${10^{-4}}$, and a batch size of 10. By choosing the number of channels in the first layer as 128, and using one attention head at resolution 16, the total number of parameters is 113,681,160 for the diffusion model and 5,452,962 for the classifier. We train the class-conditional DDPM model for 50,000 iterations and the classifier network for 20,000 iterations, which takes about one day on an NVIDIA Quadro RTX 6000 GPU. We used Pytorch 1.7.1 as software framework. The CheXpert and the BRATS2020 dataset are used for the evaluation of our method.
	
	\subsubsection{CheXpert}
	This dataset contains lung X-ray images. For training, we choose 14,179 subjects of the healthy control group, as well as 16,776 subjects suffering from pleural effusions. The images are of size $ 256 \times 256$ and normalized to values between $0$ and $1$. The test set comprises 200 images of each class.
	\subsubsection{BRATS2020}
	This dataset contains 3D brain Magnetic Resonance (MR) images of subjects with a brain tumor, as well as pixel-wise  ground truth labels. Every subject is scanned with four different MR sequences, namely, T1-weighted, T2-weighted, FLAIR, and T1-weighted with contrast enhancement. Since we focus on a 2D approach, we only consider axial slices. Each slice contains the aforementioned four channels, is padded to a size of $256 \times 256$, and normalized to values between 0 and 1. Since tumors mostly occur in the middle of the brain, we exclude the lowest 80 slices and the uppermost 26 slices. A slice is considered healthy if no tumor is found on the ground truth label mask. All other slices get the image-level label \textit{diseased}. Our training set includes 5,598 healthy slices, and 10,607 diseased slices. The test set consists of 1,082 slices containing a tumor, and 705 slices without.

	\begin{figure}[h]
		\centering
		\resizebox{1\textwidth}{!}{
			\begin{tikzpicture}
			\node[draw=black, inner sep=0pt, thick] at (0, 0) {\includegraphics[scale=0.2]{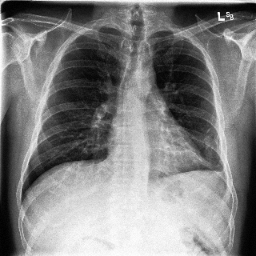}};
			\node[draw=black, inner sep=0pt, thick] at (0, 1.8) {\includegraphics[scale=0.2]{chexpert/05976/org.png}};
			\node[draw=black, inner sep=0pt, thick] at (0, 3.6) {\includegraphics[scale=0.2]{chexpert/05976/org.png}};
			\node[draw=black, inner sep=0pt, thick] at (0, 5.4) {\includegraphics[scale=0.2]{chexpert/05976/org.png}};
			\node[draw=black, inner sep=0pt, thick] at (6, 0) {\includegraphics[scale=0.2]{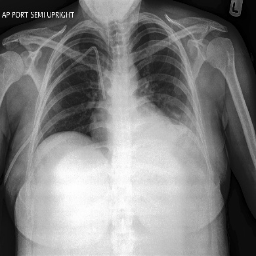}};
			\node[draw=black, inner sep=0pt, thick] at (6, 1.8) {\includegraphics[scale=0.2]{chexpert/05883/org.png}};
			\node[draw=black, inner sep=0pt, thick] at (6, 3.6) {\includegraphics[scale=0.2]{chexpert/05883/org.png}};
			\node[draw=black, inner sep=0pt, thick] at (6, 5.4) {\includegraphics[scale=0.2]{chexpert/05883/org.png}};
			\node[] at (0,6.5)  {\scriptsize Input};
			\node[] at (6,6.5)  {\scriptsize Input};
			\node[] at (1.8,6.5)  {\scriptsize Output};
			\node[] at (7.8,6.5)  {\scriptsize  Output};
			\node[] at (3.6,6.5)  {\scriptsize Anomaly Map};
			\node[] at (9.6,6.5)  {\scriptsize Anomaly Map};
			\node[draw=black, inner sep=0pt, thick] at (1.8, 5.4) {\includegraphics[scale=0.2]{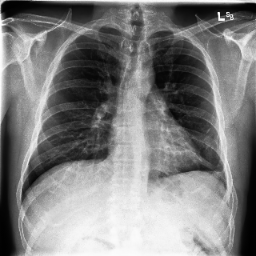}};
			\node[draw=black, inner sep=0pt, thick] at (3.6,5.4) {\includegraphics[scale=0.175]{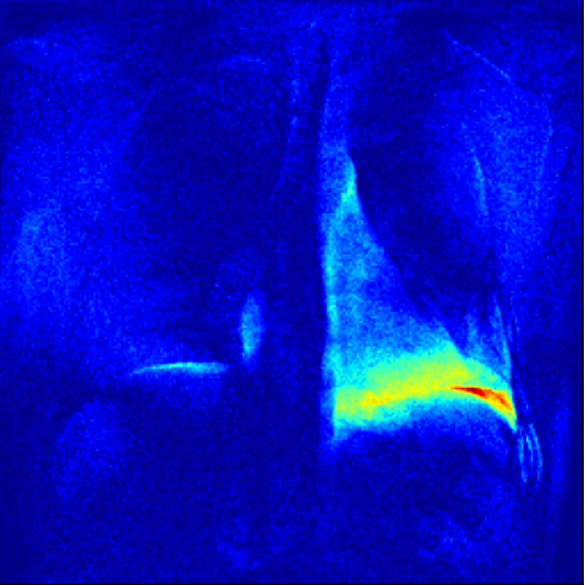}};
			\node[draw=black, inner sep=0pt, thick] at (1.8,3.6)
			{\includegraphics[scale=0.2]{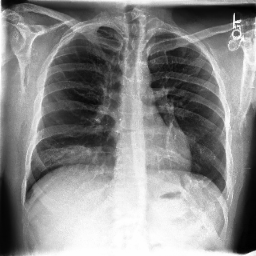}};
			\node[draw=black, inner sep=0pt, thick] at (3.6,3.6) {\includegraphics[scale=0.24]{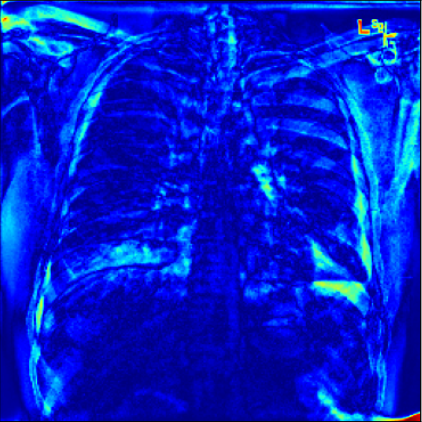}};
			\node[draw=black, inner sep=0pt, thick] at (1.8,1.8)
			{\includegraphics[scale=0.2]{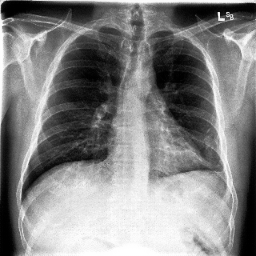}};
			\node[draw=black, inner sep=0pt, thick] at (3.6,1.8) {\includegraphics[scale=0.238]{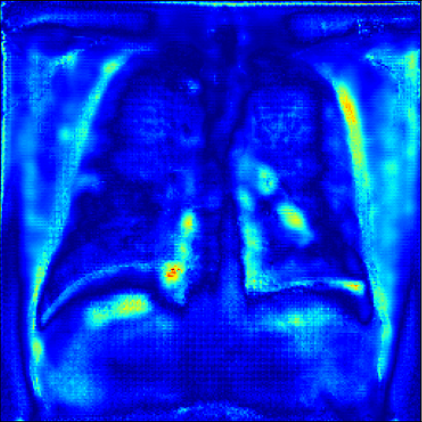}};
			\node[draw=black, inner sep=0pt, thick] at (1.8,0)
			{\includegraphics[scale=0.2]{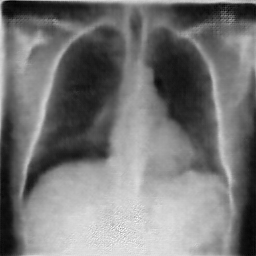}};
			\node[draw=black, inner sep=0pt, thick] at (3.6,0) {\includegraphics[scale=0.24]{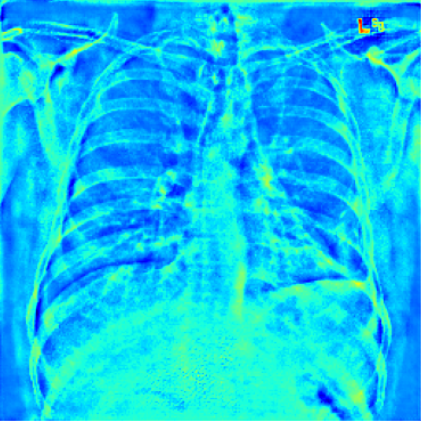}};
			\node[draw=black, inner sep=0pt, thick] at (7.8, 5.4) {\includegraphics[scale=0.2]{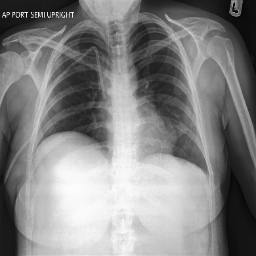}};
			\node[draw=black, inner sep=0pt, thick] at (9.6,5.4) {\includegraphics[scale=0.139]{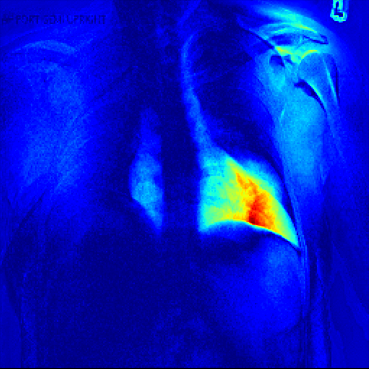}};
			\node[draw=black, inner sep=0pt, thick] at (7.8,3.6)
			{\includegraphics[scale=0.2]{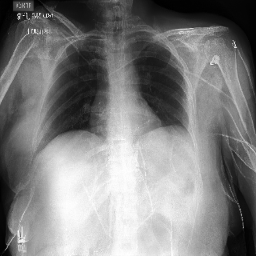}};
			\node[draw=black, inner sep=0pt, thick] at (9.6,3.6) {\includegraphics[scale=0.14]{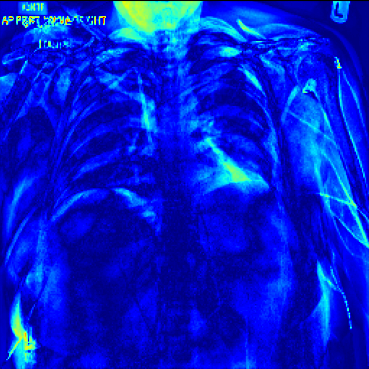}};
			\node[draw=black, inner sep=0pt, thick] at (7.8,1.8)
			{\includegraphics[scale=0.2]{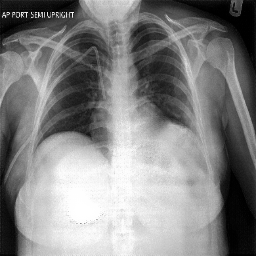}};
			\node[draw=black, inner sep=0pt, thick] at (9.6,1.8) {\includegraphics[scale=0.14]{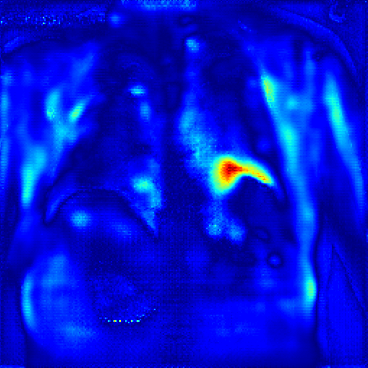}};
			\node[draw=black, inner sep=0pt, thick] at (7.8,0)
			{\includegraphics[scale=0.2]{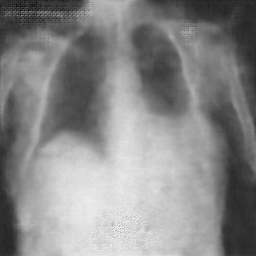}};
			\node[draw=black, inner sep=0pt, thick] at (9.6,0) {\includegraphics[scale=0.14]{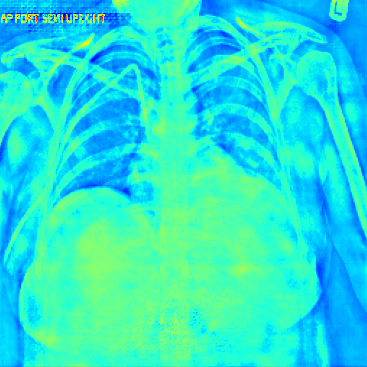}};
			\node[rotate=90] at (-1.2,5.4) {\scriptsize\textbf{Ours}};
			\node[rotate=90] at (-1.2,3.6) {\scriptsize DDPM};
			\node[rotate=90] at (-1.2,1.8) {\scriptsize FP-GAN};
			\node[rotate=90] at (-1.2,0) {\scriptsize VAE};
			\end{tikzpicture}}
		\caption{\label{chexpert1}Results for two X-ray images of the CheXpert dataset for $L=500$ and $s=100$.}
	\end{figure}
	
	\section{Results and Discussion}
\begin{figure}
		\centering
		\resizebox{0.92\textwidth}{!}{
			\begin{tikzpicture}
			\node[draw=black, inner sep=0pt, thick] at (0, 2.2) {\includegraphics[scale=0.25]{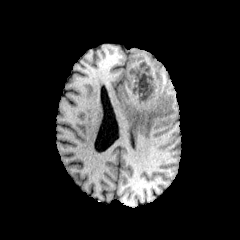}};
			\node[] at (0,3.5)  {\scriptsize T1};
			\node[draw=black, inner sep=0pt, thick] at (2.2, 2.2) {\includegraphics[scale=0.25]{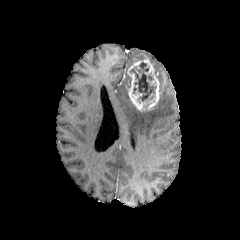}};
			\node[] at (2.2, 3.5)  {\scriptsize T1ce};
			\node[draw=black, inner sep=0pt, thick] at (4.4, 2.2) {\includegraphics[scale=0.25]{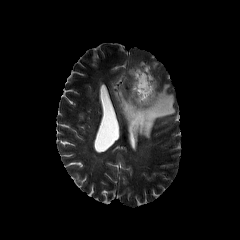}};
			\node[] at (4.4, 3.5) {\scriptsize T2};
			\node[draw=black, inner sep=0pt, thick] at (6.6, 2.2) {\includegraphics[scale=0.25]{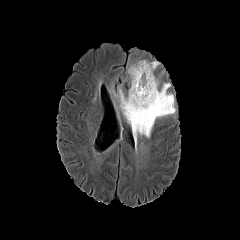}};
			\node[] at (6.6, 3.5) {\scriptsize FLAIR};
			\node[draw=white, inner sep=0pt, thick] at (8.8,2.2) {\includegraphics[scale=0.255]{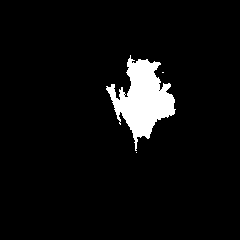}};
			\node[] at (8.8, 3.5) {\scriptsize Anomaly Map};
			\node[draw=black, inner sep=0pt, thick] at (0, 0) {\includegraphics[scale=0.25]{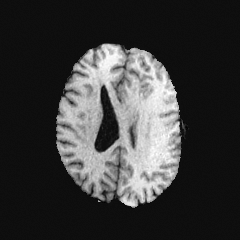}};
			\node[draw=black, inner sep=0pt, thick] at (2.2, 0) {\includegraphics[scale=0.25]{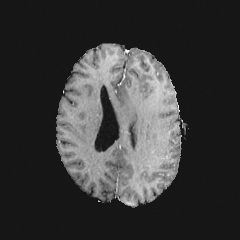}};
			\node[draw=black, inner sep=0pt, thick] at (4.4,  0) {\includegraphics[scale=0.25]{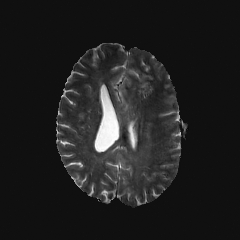}};
			\node[draw=black, inner sep=0pt, thick] at (6.6,  0) {\includegraphics[scale=0.25]{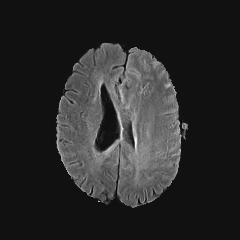}};
			\node[draw=white, inner sep=0pt, thick] at (8.8,0)  {\includegraphics[scale=0.166]{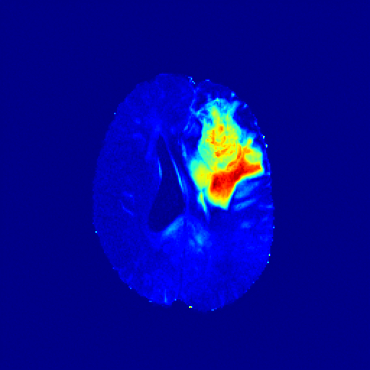}};
			\node[draw=black, inner sep=0pt, thick] at (0, -2.2) {\includegraphics[scale=0.235]{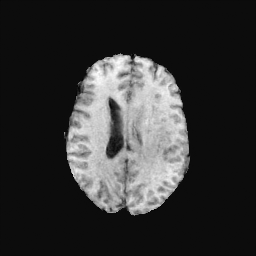}};
			\node[draw=black, inner sep=0pt, thick] at (2.2, -2.2) {\includegraphics[scale=0.235]{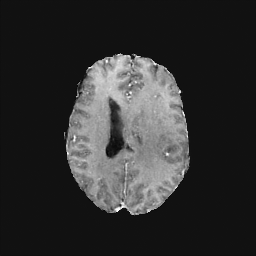}};
			\node[draw=black, inner sep=0pt, thick] at (4.4,  -2.2) {\includegraphics[scale=0.235]{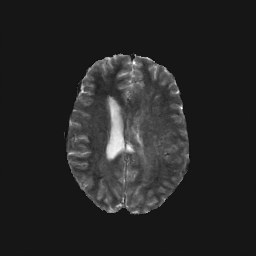}};
			\node[draw=black, inner sep=0pt, thick] at (6.6,  -2.2) {\includegraphics[scale=0.235]{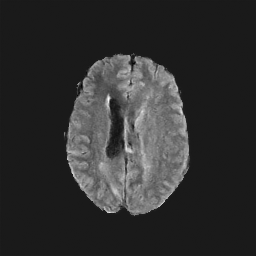}};
			\node[draw=white, inner sep=0pt, thick] at (8.8,-2.2)  {\includegraphics[scale=0.166]{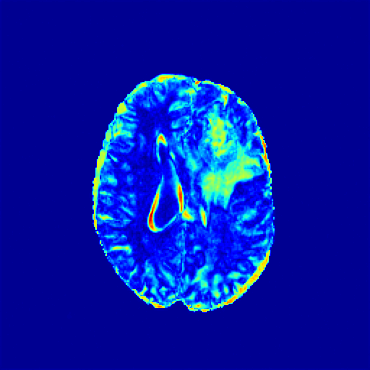}};
			\node[draw=black, inner sep=0pt, thick] at (0, -4.4) {\includegraphics[scale=0.235]{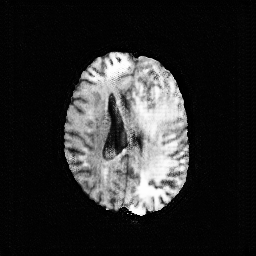}};
			\node[draw=black, inner sep=0pt, thick] at (2.2, -4.4) {\includegraphics[scale=0.235]{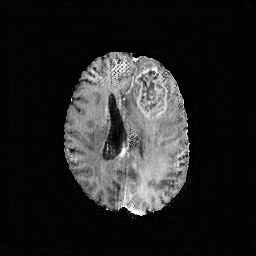}};
			\node[draw=black, inner sep=0pt, thick] at (4.4,  -4.4) {\includegraphics[scale=0.235]{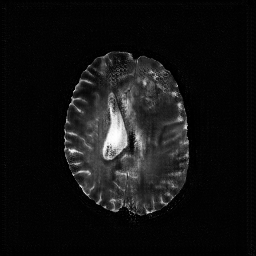}};
			\node[draw=black, inner sep=0pt, thick] at (6.6,  -4.4) {\includegraphics[scale=0.235]{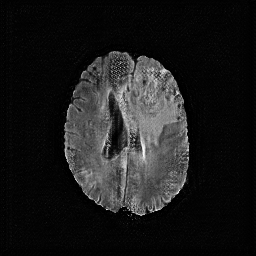}};
			\node[draw=white, inner sep=0pt, thick] at (8.8,-4.4)  {\includegraphics[scale=0.166]{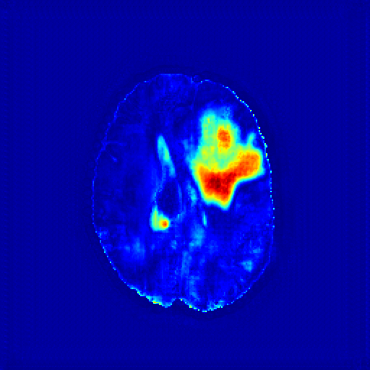}};
			\node[draw=black, inner sep=0pt, thick] at (0, -6.6) {\includegraphics[scale=0.25]{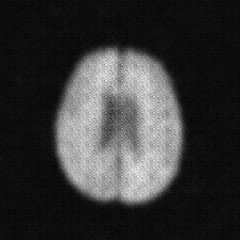}};
			\node[draw=black, inner sep=0pt, thick] at (2.2,-6.6) {\includegraphics[scale=0.25]{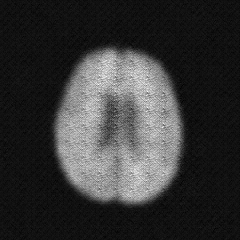}};
			\node[draw=black, inner sep=0pt, thick] at (4.4, -6.6) {\includegraphics[scale=0.25]{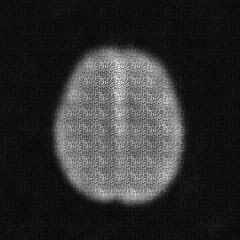}};
			\node[draw=black, inner sep=0pt, thick] at (6.6,  -6.6) {\includegraphics[scale=0.25]{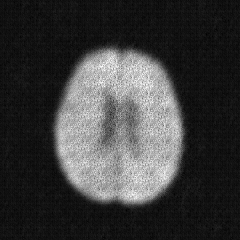}};
			\node[draw=white, inner sep=0pt, thick] at (8.8,-6.6)  {\includegraphics[scale=0.166]{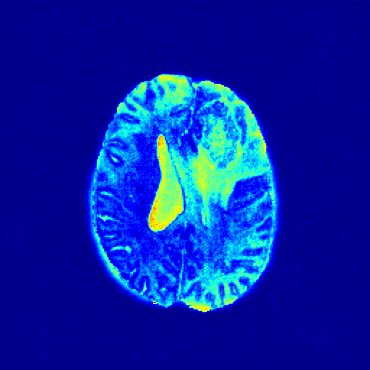}};
			\node[rotate=90]  at (-1.3,0) {\scriptsize \textbf{Ours}};
			\node[rotate=90] at (-1.3,2.2) {\scriptsize Input};
			\node[rotate=90] at (-1.3,-2.2) {\scriptsize DDPM};
			\node[rotate=90] at (-1.3,-4.4) {\scriptsize FP-GAN};
			\node[rotate=90] at (-1.3,-6.6) {\scriptsize VAE};
			\end{tikzpicture}}
		\caption{Results for an image of the BRATS2020 dataset for ${L=500}$ and ${s=100}$.} \label{brats1}
	\end{figure}
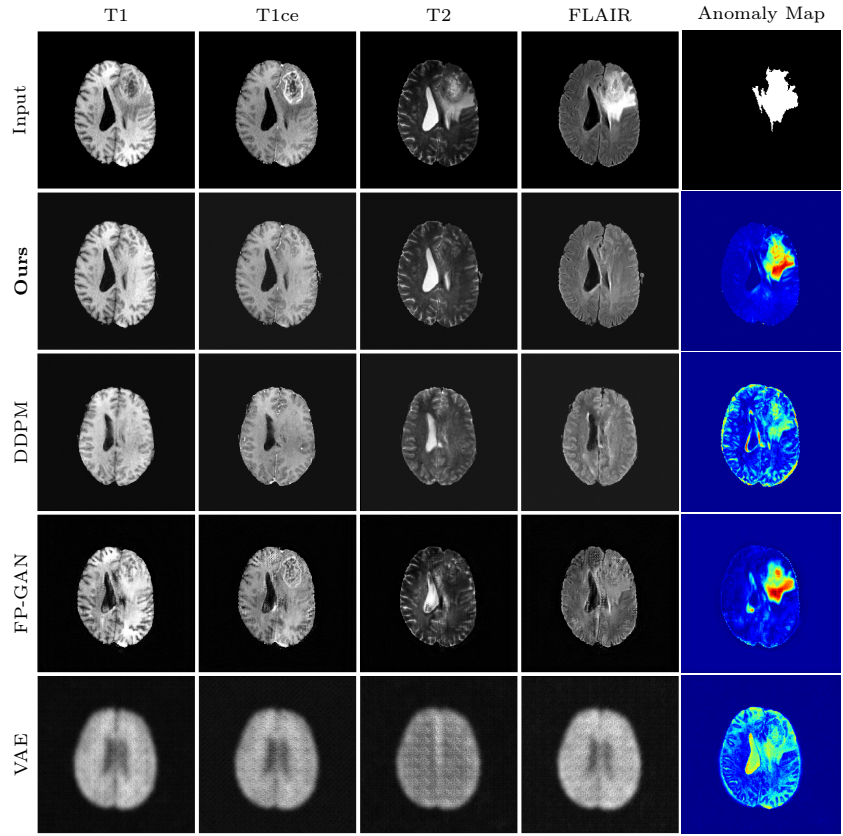

	For the evaluation of our method, we compare our method to the Fixed-Point GAN (FP-GAN) \cite{fixed}, and the variational autoencoder (VAE) proposed in \cite{autobrats2}. 
	As an ablation study, we add random noise for L steps to the input image using and perform the sampling using the DDPM sampling scheme with classifier guidance. In all experiments, we set ${s=100}$ and ${L=500}$. 
	In Figure~\ref{chexpert1}, we show two exemplary patient images of the CheXpert dataset, and apply all comparing methods to generate the corresponding healthy image. We observe that compared to the other methods, our approach generates realistic looking images and preserves all the details of the input image, which leads to a very detailed anomaly map. The other methods either change other parts image, or are not able to find an anomaly.
	Figure~\ref{brats1} shows the results for all four MR sequences for an exemplary image of the BRATS2020 dataset. 
	More examples can be found in the supplementary material. 
	Of all methods, only the VAE tries to reconstruct the right ventricle.  
	Comparing our results to the results of DDPM, we see that encoding information in noise using the deterministic noising process of DDIM brings the advantage that all details of the input image can be reconstructed. In contrast, we see that sampling with the DDPM approach changes the
	basic anatomy of the input image. The computation of a complete image translation takes about 158s. This longish running time is mainly due to the iterative image generation process.  We could speed up this process by choosing a smaller $L$, or by skipping timesteps in the DDIM sampling scheme. However, we observed that this degrades the image quality.\\
	In \cite{challenging}, using the reconstruction error as anomaly score has received some criticism. It was shown that a simple method based on histogram equalization could outperform neural networks and state that reconstruction quality does not correlate well with the Dice score. As an alternative, anomaly scores of other types of methods, i.e., the log-likelihood of density estimation models, will be explored in future work.

	\subsubsection{Hyperparameter Sensitivity}
\begin{figure}[h]
		\centering
		\begin{minipage}[t]{0.45\textwidth}
			{\resizebox{0.9\textwidth}{!}{
\begin{tikzpicture}

\definecolor{color0}{rgb}{0,0.75,0.75}

\begin{axis}[
legend cell align={left},
legend style={fill opacity=0.8, draw opacity=1, text opacity=1, at={(0.91,0.5)}, anchor=east, draw=white!80!black},
tick align=outside,
tick pos=left,
title={\large \textbf{Dice Score}},
x grid style={white!69.0196078431373!black},
xlabel={gradient scale},
xmin=-32.25, xmax=787.25,
xtick style={color=black},
y grid style={white!69.0196078431373!black},
ylabel={avg Dice score},
ymin=0.18304770886898, ymax=0.722336533665657,
ytick style={color=black},
ytick={0.1,0.2,0.3,0.4,0.5,0.6,0.7,0.8},
yticklabels={0.1,0.2,0.3,0.4,0.5,0.6,0.7,0.8}
]
\addplot [semithick, blue, mark=*, mark size=3, mark options={solid}]
table {%
5 0.647607564926147
10 0.684657752513885
20 0.697823405265808
50 0.697234869003296
100 0.684662461280823
250 0.647283792495728
350 0.627332925796509
500 0.603253126144409
750 0.569113552570343
};
\addlegendentry{L=500}
\addplot [semithick, red, mark=*, mark size=3, mark options={solid}]
table {%
5 0.277640402317047
10 0.343390882015228
20 0.476263910531998
50 0.612984418869019
100 0.642866969108582
250 0.662366569042206
500 0.657480001449585
750 0.654248356819153
};
\addlegendentry{L=250}
\addplot [semithick, black, mark=*, mark size=3, mark options={solid}]
table {%
5 0.428322196006775
10 0.371534168720245
20 0.318564176559448
50 0.265155434608459
100 0.23495227098465
250 0.214335948228836
500 0.207560837268829
750 0.212064072489738
};
\addlegendentry{L=750}
\addplot [semithick, green!50!black, dashed]
table {%
5 0.693
10 0.693
20 0.693
50 0.693
100 0.693
250 0.693
350 0.693
500 0.693
750 0.693
};
\addlegendentry{FP-GAN}
\addplot [semithick, color0, dashed]
table {%
5 0.239
10 0.239
20 0.239
50 0.239
100 0.239
250 0.239
350 0.239
500 0.239
750 0.239
};
\addlegendentry{VAE}
\end{axis}

\end{tikzpicture}}}
		\end{minipage}
		\qquad
		\begin{minipage}[t]{0.45\textwidth}
			{\resizebox{0.9\textwidth}{!}{
\begin{tikzpicture}

\definecolor{color0}{rgb}{0,0.75,0.75}

\begin{axis}[
legend cell align={left},
legend style={fill opacity=0.8, draw opacity=1, text opacity=1, at={(0.91,0.5)}, anchor=east, draw=white!80!black},
tick align=outside,
tick pos=left,
title={\large \textbf{AUROC}},
x grid style={white!69.0196078431373!black},
xlabel={gradient scale},
xmin=-32.25, xmax=787.25,
xtick style={color=black},
y grid style={white!69.0196078431373!black},
ylabel={avg AUROC},
ymin=0.910819466429275, ymax=0.990545201872774,
ytick style={color=black},
ytick={0.91,0.92,0.93,0.94,0.95,0.96,0.97,0.98,0.99,1},
yticklabels={0.91,0.92,0.93,0.94,0.95,0.96,0.97,0.98,0.99,1.00}
]
\addplot [semithick, blue, mark=*, mark size=3, mark options={solid}]
table {%
5 0.980395209479452
10 0.984832099477862
20 0.986027962977908
50 0.986921304807161
100 0.986654321033975
250 0.985492409084434
350 0.984593787390702
500 0.983212486454774
750 0.981348523835677
};
\addlegendentry{L=500}
\addplot [semithick, red, mark=*, mark size=3, mark options={solid}]
table {%
5 0.9341085382295
10 0.951451561706018
20 0.966907618542192
50 0.979228815273021
100 0.982699119742456
250 0.984363445761624
500 0.983443053772777
750 0.983081903789456
};
\addlegendentry{L=250}
\addplot [semithick, black, mark=*, mark size=3, mark options={solid}]
table {%
5 0.963188313854861
10 0.95424934722076
20 0.944176530285981
50 0.929673087283602
100 0.919575126887924
250 0.914443363494889
500 0.91721953508827
750 0.922768287915514
};
\addlegendentry{L=750}
\addplot [semithick, green!50!black, dashed]
table {%
5 0.965
10 0.965
20 0.965
50 0.965
100 0.965
250 0.965
350 0.965
500 0.965
750 0.965
};
\addlegendentry{FP-GAN}
\addplot [semithick, color0, dashed]
table {%
5 0.944
10 0.944
20 0.944
50 0.944
100 0.944
250 0.944
350 0.944
500 0.944
750 0.944
};
\addlegendentry{VAE}
\end{axis}

\end{tikzpicture}}}
		\end{minipage}
		\label{graphs}
		\caption{Average Dice and AUROC scores on the  test set for different $s$ and $L$.} 
	\end{figure}
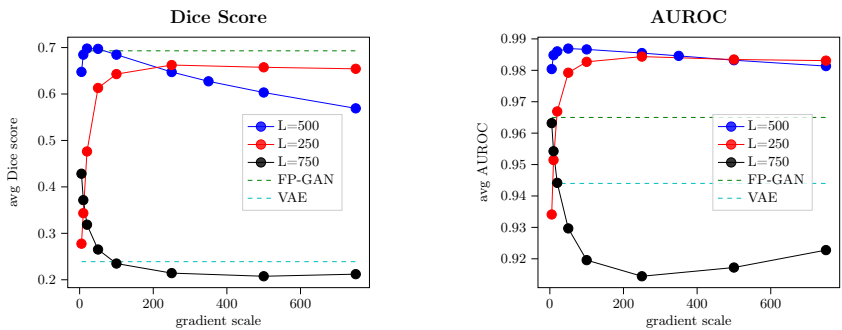

	Our method has two major hyperparameters, the classifier gradient scale $s$ and the noise level $L$. We performed experiments to evaluate the sensitivity of our method to changes of $s$ and $L$. On the BRATS2020 dataset, we have pixel-wise ground truth labels, which enable us to calculate the Dice score and the Area under the receiver operating statistics (AUROC) for diseased slices. For the Dice score, we use the average Otsu thresholding \cite{otsu} on the anomaly maps. In Figure~\ref{graphs}, we show the average Dice and  AUROC scores on the test set  with respect to the gradient scale $s$ for \mbox{different noise levels $L$.} The scores for the comparing methods FP-GAN and VAE are shown in horizontal bars in Figure~\ref{graphs}.\\
	Figure \ref{series_s} shows an exemplary FLAIR image. We fix ${L = 500}$ and show the sampled results for various values of $s$. If we choose $s$ too small, the tumor cannot be removed. 
	However, if we choose $s$ too large, additional artefacts are introduced to the image. Those artefacts are mainly at the border of the brain, and lead to a decrease in the Dice score. In Figure~\ref{series_L}, we fix ${s = 100}$, and show the sampled results for the same image for varying noise levels $L$. If $L$ is chosen too large, this results in a destruction of the images. If $L$ is chosen too small, the model does not have enough freedom to remove the tumor from the image.
	\begin{figure}
		\centering
		\resizebox{0.95\textwidth}{!}{
			\begin{tikzpicture}
			\node[draw=black, inner sep=0pt, thick] at (0, 2) {\includegraphics[scale=0.23]{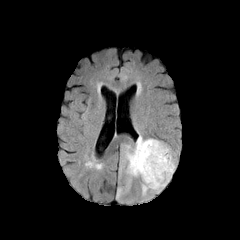}};
			\node[] at (0,3.2)  {\scriptsize Input};
			\node[draw=black, inner sep=0pt, thick] at (2, 2) {\includegraphics[scale=0.23]{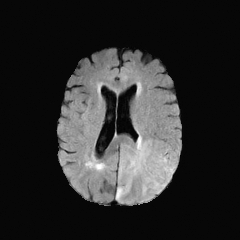}};
			\node[] at (2, 3.2)  {\scriptsize s=10};
			\node[draw=black, inner sep=0pt, thick] at (4, 2){\includegraphics[scale=0.23]{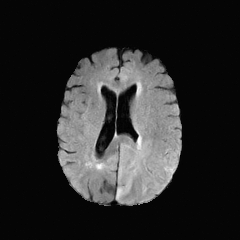}};
			\node[] at (4, 3.2) {\scriptsize s=250};
			\node[draw=black, inner sep=0pt, thick] at (6, 2){\includegraphics[scale=0.23]{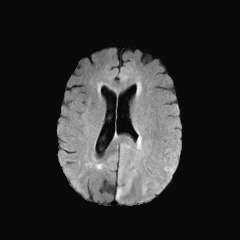}};
			\node[] at (6, 3.2) {\scriptsize s=500};
			\node[draw=black, inner sep=0pt, thick] at (8, 2) {\includegraphics[scale=0.23]{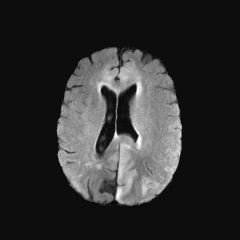}};
			\node[] at (8, 3.2) {\scriptsize s=1000};
			\node[draw=white, inner sep=0pt, thick] at (10,2) {\includegraphics[scale=0.23]{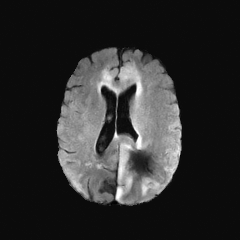}};
			\node[] at (10, 3.2) {\scriptsize s=1500};
			\node[draw=black, inner sep=0pt, thick] at (0, 0) {\includegraphics[scale=0.23]{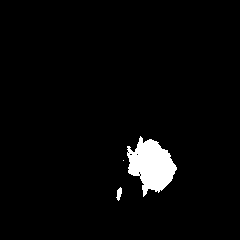}};
			\node[draw=white, inner sep=0pt, thick] at (2, 0) {\includegraphics[scale=0.152]{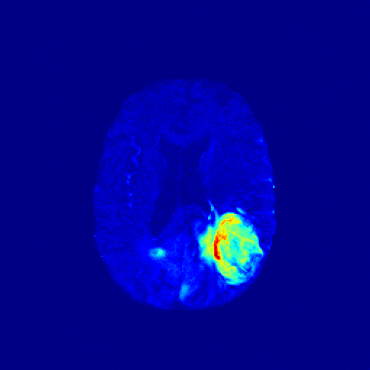}};
			\node[draw=white, inner sep=0pt, thick] at (4,  0) {\includegraphics[scale=0.152]{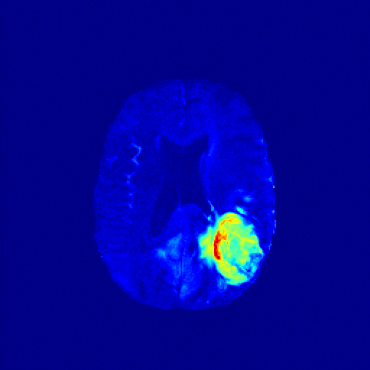}};
			\node[draw=white, inner sep=0pt, thick] at (6,  0) {\includegraphics[scale=0.152]{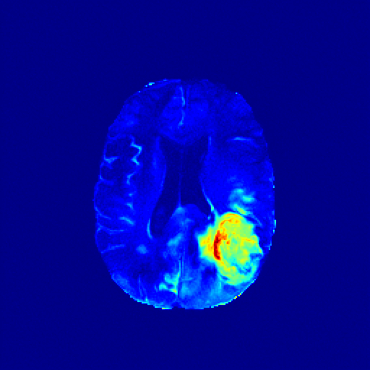}};
			
			\node[draw=white, inner sep=0pt, thick] at (8,  0) {\includegraphics[scale=0.152]{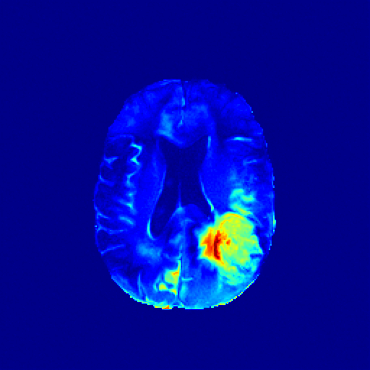}};
			\node[draw=white, inner sep=0pt, thick] at (10,0)  {\includegraphics[scale=0.152]{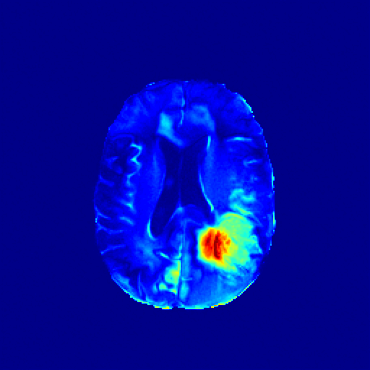}};
			\node[rotate=90]  at (-1.2,0) {\scriptsize Anomaly Map};
			\node[rotate=90] at (-1.2,2) {\scriptsize Images};
			\end{tikzpicture}}
		\caption{Illustration of the effect of the gradient scale $s$ for a fixed noise level $L=500$.} \label{series_s}
	\end{figure}
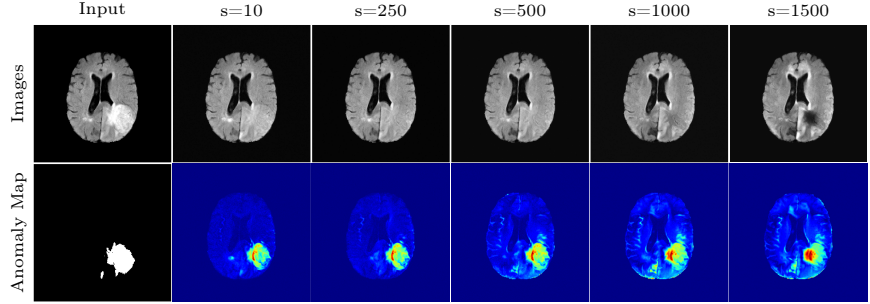
	\begin{figure}[h]
		\centering
		\resizebox{0.95\textwidth}{!}{
			\begin{tikzpicture}
			\node[draw=white, inner sep=0pt, thick] at (0, 2) {\includegraphics[scale=0.22]{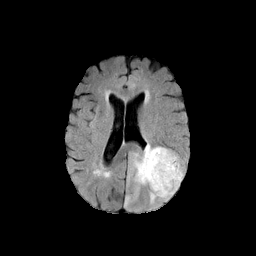}};
			\node[] at (0,3.2)  {\scriptsize Input};
			\node[draw=white, inner sep=0pt, thick] at (2, 2) {\includegraphics[scale=0.22]{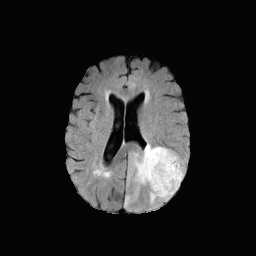}};
			\node[] at (2, 3.2)  {\scriptsize L=100};
			\node[draw=white, inner sep=0pt, thick] at (4, 2){\includegraphics[scale=0.22]{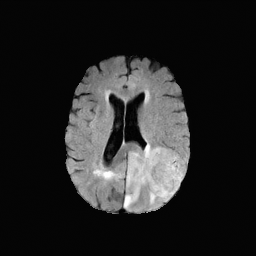}};
			\node[draw=white, inner sep=0pt, thick] at (6, 2){\includegraphics[scale=0.22]{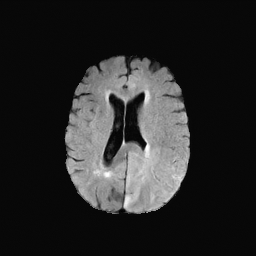}};
			\node[] at (4, 3.2) {\scriptsize L=250};
			\node[draw=white, inner sep=0pt, thick] at (8, 2) {\includegraphics[scale=0.22]{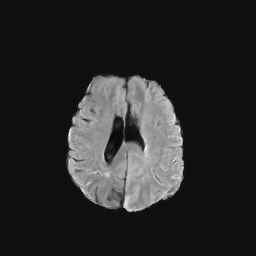}};
			\node[] at (6, 3.2) {\scriptsize L=500};
			\node[draw=white, inner sep=0pt, thick] at (10,2) {\includegraphics[scale=0.22]{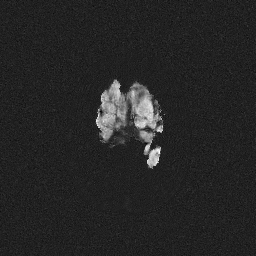}};
			\node[] at (8, 3.2) {\scriptsize L=750};
			\node[] at (10, 3.2) {\scriptsize L=1000};
			\node[draw=white, inner sep=0pt, thick] at (0, 0) {\includegraphics[scale=0.235]{brats/imageseries/003487/gt.png}};
			\node[draw=white, inner sep=0pt, thick] at (2, 0) {\includegraphics[scale=0.152]{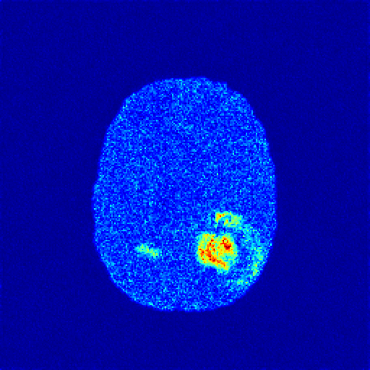}};
			\node[draw=white, inner sep=0pt, thick] at (4, 0) {\includegraphics[scale=0.152]{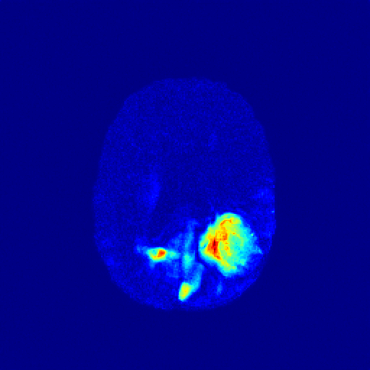}};
			\node[draw=white, inner sep=0pt, thick] at (6,  0) {\includegraphics[scale=0.152]{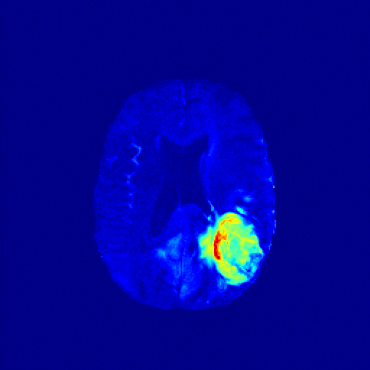}};
			\node[draw=white, inner sep=0pt, thick] at (8,  0) {\includegraphics[scale=0.152]{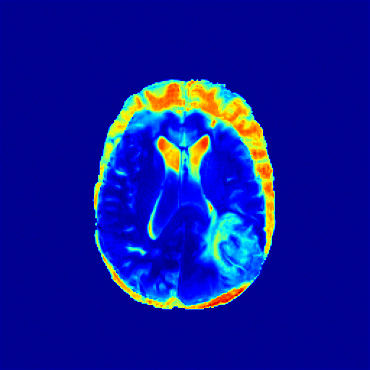}};
			\node[draw=white, inner sep=0pt, thick] at (10,0)  {\includegraphics[scale=0.152]{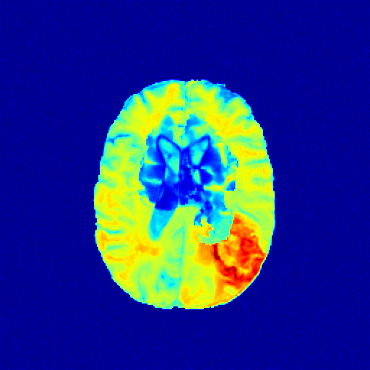}};
			\node[rotate=90]  at (-1.2,0) {\scriptsize Anomaly Map};
			\node[rotate=90] at (-1.2,2) {\scriptsize Images};
			\end{tikzpicture}}
		\caption{Illustration of the effect of the noise level $L$, for a fixed gradient scale $s=100$.} \label{series_L}
	\end{figure}
	\subsubsection{Translation of a Healthy Subject}
\begin{figure}
		\centering
		\resizebox{0.85\textwidth}{!}{
			\begin{tikzpicture}
			\node[draw=white, inner sep=0pt, thick] at (0, 2) {\includegraphics[scale=0.22]{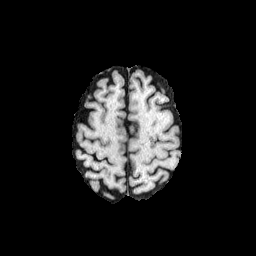}};
			\node[] at (0,3.2)  {\scriptsize T1};
			\node[draw=white, inner sep=0pt, thick] at (2, 2) {\includegraphics[scale=0.22]{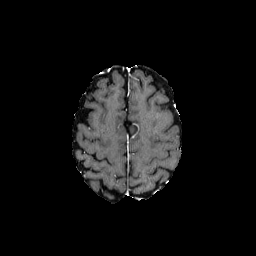}};
			\node[] at (2, 3.2)  {\scriptsize T1ce};
			\node[draw=white, inner sep=0pt, thick] at (4, 2){\includegraphics[scale=0.22]{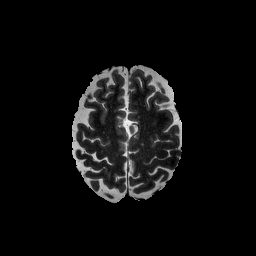}};
			\node[] at (4, 3.2) {\scriptsize T2};
			\node[draw=white, inner sep=0pt, thick] at (6, 2){\includegraphics[scale=0.22]{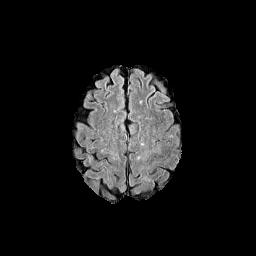}};
			\node[] at (6, 3.2) {\scriptsize FLAIR};
			\node[draw=white, inner sep=0pt, thick] at (8, 2) {\includegraphics[scale=0.235]{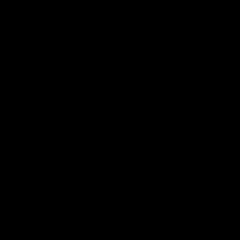}};
			\node[] at (8, 3.2) {\scriptsize Anomaly Map};
			\node[draw=white, inner sep=0pt, thick] at (0, 0) {\includegraphics[scale=0.22]{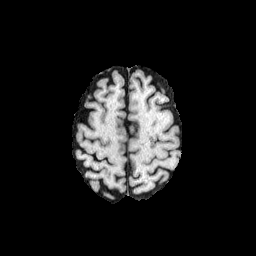}};
			\node[draw=white, inner sep=0pt, thick] at (2, 0) {\includegraphics[scale=0.22]{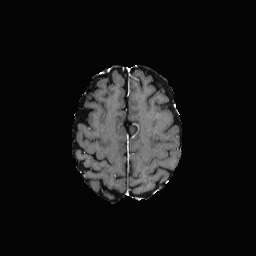}};
			\node[draw=white, inner sep=0pt, thick] at (4,  0) {\includegraphics[scale=0.22]{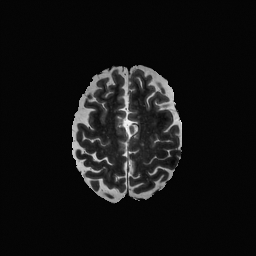}};
			\node[draw=white, inner sep=0pt, thick] at (6,  0) {\includegraphics[scale=0.22]{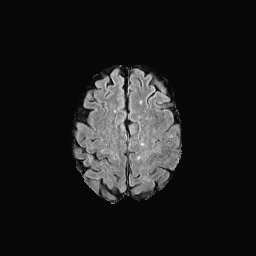}};
			
			\node[draw=white, inner sep=0pt, thick] at (8,  0) {\includegraphics[scale=0.2]{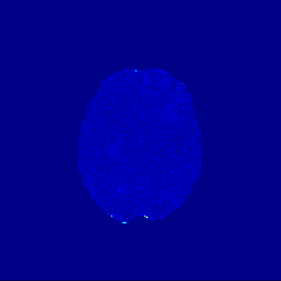}};
			
			\node[rotate=90]  at (-1.2,0) {\scriptsize\textbf{Ours}};
			\node[rotate=90] at (-1.2,2) {\scriptsize Input};
			\end{tikzpicture}}
		\caption{Results of the presented method for an image without a tumor. The difference between the input image and the synthetic image is close to zero. } \label{healthy}
	\end{figure}
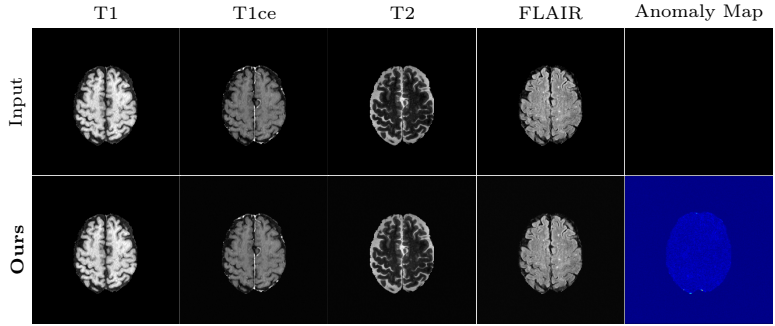

	If an input image shows a healthy subject, our method should not make any changes to this image. In Figure~ \ref{healthy}, we evaluate our approach on a healthy slice of the BRATS dataset. We get a very detailed reconstruction of the image, resulting in an anomaly map close to zero.
	
	\section{Conclusion}
	In this paper, we presented a novel weakly supervised anomaly detection method by combining the iterative DDIM noising and denoising schemes, and classifier guidance. No changes were made to the loss function or the training scheme of the original implementations, making the training on other datasets straightforward. We applied our method for anomaly detection on two different medical datasets and successfully translated images of patients to images without pathologies. Our method only performs changes in the anomalous regions of the image to achieve the translation to a healthy subject. This improves the quality of the anomaly maps. We point out that we achieve a detail-consistent image-to-image translation without the need of changing the architecture or training procedure. We achieve excellent results on the BRATS2020 and the CheXpert dataset.
	
	\subsubsection{Acknowledgements} This research was supported by the Novartis FreeNovation initiative and the Uniscientia
	Foundation (project $\#$147-2018).
	
	%
	
	\bibliographystyle{splncs04}
	\bibliography{adddim}

\begin{thebibliography}{10}
\providecommand{\url}[1]{\texttt{#1}}
\providecommand{\urlprefix}{URL }
\providecommand{\doi}[1]{https://doi.org/#1}

\bibitem{saliency}
Arun, N.T., Gaw, N., Singh, P., Chang, K., Hoebel, K.V., Patel, J., Gidwani,
  M., Kalpathy-Cramer, J.: Assessing the validity of saliency maps for
  abnormality localization in medical imaging. arXiv preprint arXiv:2006.00063
  (2020)

\bibitem{brats2}
Bakas, S., Akbari, H., Sotiras, A., Bilello, M., Rozycki, M., Kirby, J.S.,
  Freymann, J.B., Farahani, K., Davatzikos, C.: Advancing the cancer genome
  atlas glioma {MRI} collections with expert segmentation labels and radiomic
  features. Scientific data  \textbf{4}(1),  1--13 (2017)

\bibitem{brats3}
Bakas, S., Reyes, M., Jakab, A., Bauer, S., Rempfler, M., Crimi, A., Shinohara,
  R.T., Berger, C., Ha, S.M., Rozycki, M., et~al.: Identifying the best machine
  learning algorithms for brain tumor segmentation, progression assessment, and
  overall survival prediction in the {BRATS} challenge. arXiv preprint
  arXiv:1811.02629  (2018)

\bibitem{diffseg}
Baranchuk, D., Voynov, A., Rubachev, I., Khrulkov, V., Babenko, A.:
  Label-efficient semantic segmentation with diffusion models. In:
  International Conference on Learning Representations (2022)

\bibitem{vagan}
Baumgartner, C.F., Koch, L.M., Tezcan, K.C., Ang, J.X., Konukoglu, E.: Visual
  feature attribution using wasserstein gans. In: Proceedings of the IEEE
  Conference on Computer Vision and Pattern Recognition. pp. 8309--8319 (2018)

\bibitem{autobrats2}
Chen, X., Konukoglu, E.: Unsupervised detection of lesions in brain mri using
  constrained adversarial auto-encoders. arXiv preprint arXiv:1806.04972
  (2018)

\bibitem{ilvr}
Choi, J., Kim, S., Jeong, Y., Gwon, Y., Yoon, S.: Ilvr: Conditioning method for
  denoising diffusion probabilistic models. arXiv preprint arXiv:2108.02938
  (2021)

\bibitem{beatgans}
Dhariwal, P., Nichol, A.: Diffusion models beat gans on image synthesis.
  Advances in Neural Information Processing Systems  \textbf{34} (2021)

\bibitem{gan}
Goodfellow, I., Pouget-Abadie, J., Mirza, M., Xu, B., Warde-Farley, D., Ozair,
  S., Courville, A., Bengio, Y.: Generative adversarial nets. Advances in
  neural information processing systems  \textbf{27} (2014)

\bibitem{ddpm}
Ho, J., Jain, A., Abbeel, P.: Denoising diffusion probabilistic models.
  Advances in Neural Information Processing Systems  \textbf{33},  6840--6851
  (2020)

\bibitem{chexpert}
Irvin, J., Rajpurkar, P., Ko, M., Yu, Y., Ciurea-Ilcus, S., Chute, C.,
  Marklund, H., Haghgoo, B., Ball, R., Shpanskaya, K., et~al.: Chexpert: A
  large chest radiograph dataset with uncertainty labels and expert comparison.
  In: Proceedings of the AAAI conference on artificial intelligence. vol.~33,
  pp. 590--597 (2019)

\bibitem{diffusemorph}
Kim, B., Han, I., Ye, J.C.: Diffusemorph: Unsupervised deformable image
  registration along continuous trajectory using diffusion models. arXiv
  preprint arXiv:2112.05149  (2021)

\bibitem{kingma}
Kingma, D.P., Welling, M.: An introduction to variational autoencoders. arXiv
  preprint arXiv:1906.02691  (2019)

\bibitem{marimont}
Marimont, S.N., Tarroni, G.: Anomaly detection through latent space restoration
  using vector quantized variational autoencoders. In: 2021 IEEE 18th
  International Symposium on Biomedical Imaging (ISBI). pp. 1764--1767. IEEE
  (2021)

\bibitem{challenging}
Meissen, F., Kaissis, G., Rueckert, D.: Challenging current semi-supervised
  anomaly segmentation methods for brain mri. arXiv preprint arXiv:2109.06023
  (2021)

\bibitem{brats1}
Menze, B.H., Jakab, A., Bauer, S., Kalpathy-Cramer, J., Farahani, K., Kirby,
  J., Burren, Y., Porz, N., Slotboom, J., Wiest, R., et~al.: The multimodal
  brain tumor image segmentation benchmark ({BRATS}). IEEE transactions on
  medical imaging  \textbf{34}(10),  1993--2024 (2014)

\bibitem{improving}
Nichol, A.Q., Dhariwal, P.: Improved denoising diffusion probabilistic models.
  In: Proceedings of the 38th International Conference on Machine Learning.
  vol.~139, pp. 8162--8171. PMLR (2021)

\bibitem{otsu}
Otsu, N.: A threshold selection method from gray-level histograms. IEEE
  transactions on systems, man, and cybernetics  \textbf{9}(1),  62--66 (1979)

\bibitem{cam}
Panwar, H., Gupta, P., Siddiqui, M.K., Morales-Menendez, R., Bhardwaj, P.,
  Singh, V.: A deep learning and grad-cam based color visualization approach
  for fast detection of covid-19 cases using chest x-ray and ct-scan images.
  Chaos, Solitons \& Fractals  \textbf{140},  110190 (2020)

\bibitem{transformers1}
Pinaya, W.H.L., Tudosiu, P.D., Gray, R., Rees, G., Nachev, P., Ourselin, S.,
  Cardoso, M.J.: Unsupervised brain anomaly detection and segmentation with
  transformers. arXiv preprint arXiv:2102.11650  (2021)

\bibitem{transformers2}
Pirnay, J., Chai, K.: Inpainting transformer for anomaly detection. arXiv
  preprint arXiv:2104.13897  (2021)

\bibitem{palette}
Saharia, C., Chan, W., Chang, H., Lee, C.A., Ho, J., Salimans, T., Fleet, D.J.,
  Norouzi, M.: Palette: Image-to-image diffusion models. arXiv preprint
  arXiv:2111.05826  (2021)

\bibitem{unitddpm}
Sasaki, H., Willcocks, C.G., Breckon, T.P.: Unit-ddpm: Unpaired image
  translation with denoising diffusion probabilistic models. arXiv preprint
  arXiv:2104.05358  (2021)

\bibitem{fixed}
Siddiquee, M.M.R., Zhou, Z., Tajbakhsh, N., Feng, R., Gotway, M.B., Bengio, Y.,
  Liang, J.: Learning fixed points in generative adversarial networks: From
  image-to-image translation to disease detection and localization. In:
  Proceedings of the IEEE/CVF international conference on computer vision. pp.
  191--200 (2019)

\bibitem{ddim}
Song, J., Meng, C., Ermon, S.: Denoising diffusion implicit models. arXiv
  preprint arXiv:2010.02502  (2020)

\bibitem{score_based}
Song, Y., Sohl-Dickstein, J., Kingma, D.P., Kumar, A., Ermon, S., Poole, B.:
  Score-based generative modeling through stochastic differential equations.
  arXiv preprint arXiv:2011.13456  (2020)

\bibitem{descargan}
Wolleb, J., Sandk{\"u}hler, R., Cattin, P.C.: Descargan: Disease-specific
  anomaly detection with weak supervision. In: International Conference on
  Medical Image Computing and Computer-Assisted Intervention. pp. 14--24.
  Springer (2020)

\bibitem{yang2021visual}
Yang, J., Xu, R., Qi, Z., Shi, Y.: Visual anomaly detection for images: A
  survey. arXiv preprint arXiv:2109.13157  (2021)

\bibitem{auto1}
Zhou, C., Paffenroth, R.C.: Anomaly detection with robust deep autoencoders.
  In: Proceedings of the 23rd ACM SIGKDD international conference on knowledge
  discovery and data mining. pp. 665--674 (2017)

\bibitem{autobrats1}
Zimmerer, D., Kohl, S.A., Petersen, J., Isensee, F., Maier-Hein, K.H.:
  Context-encoding variational autoencoder for unsupervised anomaly detection.
  arXiv preprint arXiv:1812.05941  (2018)

\end{thebibliography}
	
\end{document}


\title{Supplementary Material}
\author{}
\institute{}
\maketitle
\section{Additional Results on the CheXpert Dataset}
\begin{figure}
\centering
\resizebox{\textwidth}{!}{
	\begin{tikzpicture}
	\node[] at (0,9)  {\scriptsize Input};
	\node[] at (2.5,9)  {\scriptsize Output};
	\node[] at (5,9)  {\scriptsize Anomaly Map};
		\node[] at (8,9)  {\scriptsize Input};
	\node[] at (10.5,9)  {\scriptsize Output};
	\node[] at (13,9)  {\scriptsize Anomaly Map};
	\node[draw=black, inner sep=0pt, thick] at (0, 2.5) {\includegraphics[scale=0.275]{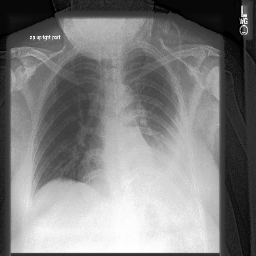}};
	\node[draw=black, inner sep=0pt, thick] at (2.5, 2.5) {\includegraphics[scale=0.275]{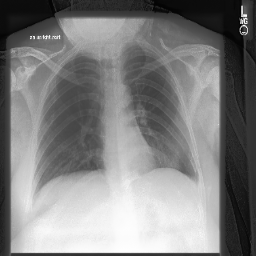}};
	\node[draw=black, inner sep=0pt, thick] at (5, 2.5) {\includegraphics[scale=0.19]{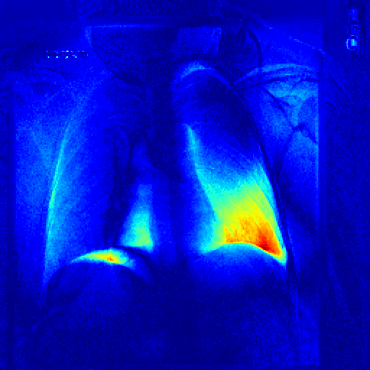}};
\node[draw=black, inner sep=0pt, thick] at (0, 5) {\includegraphics[scale=0.27]{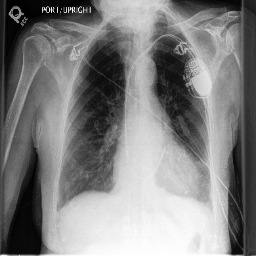}};
	\node[draw=black, inner sep=0pt, thick] at (2.5, 5) {\includegraphics[scale=0.275]{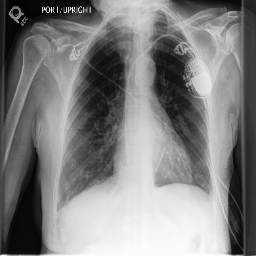}};
	\node[draw=black, inner sep=0pt, thick] at (5, 5) {\includegraphics[scale=0.19]{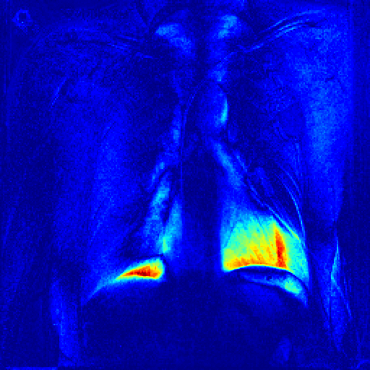}};
	\node[draw=black, inner sep=0pt, thick] at (0, 7.5) {\includegraphics[scale=0.275]{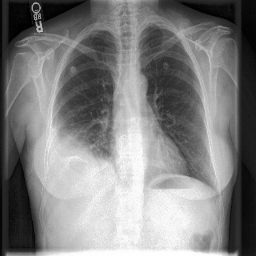}};
	\node[draw=black, inner sep=0pt, thick] at (2.5, 7.5) {\includegraphics[scale=0.275]{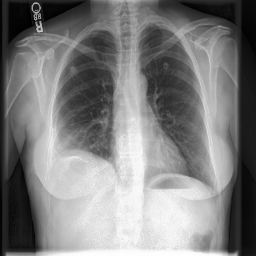}};
	\node[draw=black, inner sep=0pt, thick] at (5, 7.5) {\includegraphics[scale=0.19]{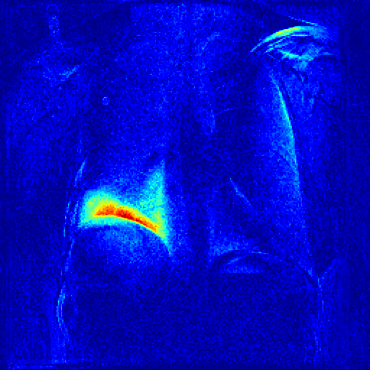}};
\node[draw=black, inner sep=0pt, thick] at (8, 7.5) {\includegraphics[scale=0.275]{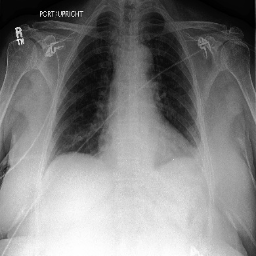}};
	\node[draw=black, inner sep=0pt, thick] at (10.5, 7.5) {\includegraphics[scale=0.275]{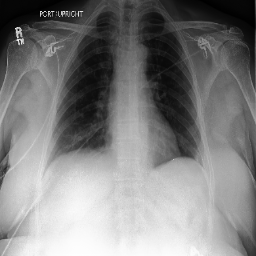}};
	\node[draw=black, inner sep=0pt, thick] at (13, 7.5) {\includegraphics[scale=0.19]{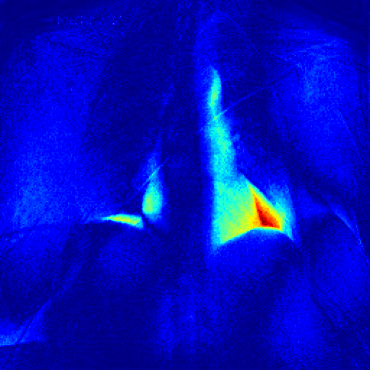}};
	\node[draw=black, inner sep=0pt, thick] at (8, 5) {\includegraphics[scale=0.275]{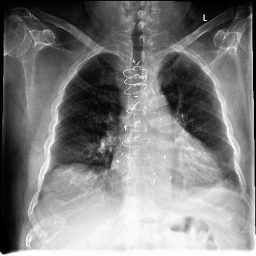}};
	\node[draw=black, inner sep=0pt, thick] at (10.5, 5) {\includegraphics[scale=0.275]{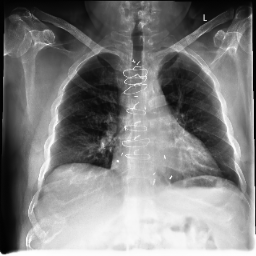}};
	\node[draw=black, inner sep=0pt, thick] at (13, 5) {\includegraphics[scale=0.19]{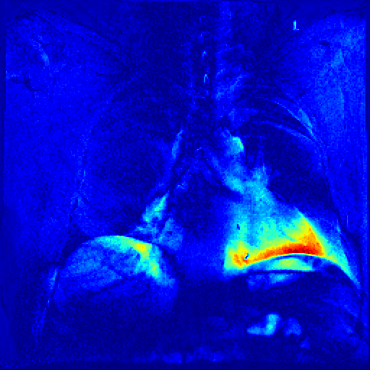}};
	\node[draw=black, inner sep=0pt, thick] at (8, 2.5) {\includegraphics[scale=0.275]{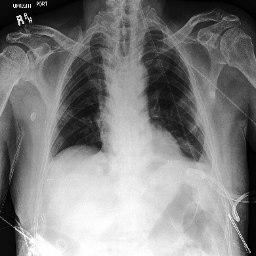}};
	\node[draw=black, inner sep=0pt, thick] at (10.5, 2.5) {\includegraphics[scale=0.275]{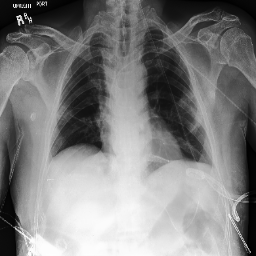}};
	\node[draw=black, inner sep=0pt, thick] at (13, 2.5) {\includegraphics[scale=0.19]{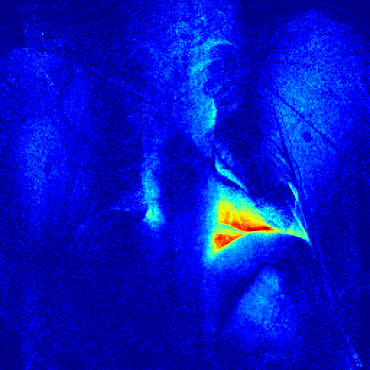}};

\end{tikzpicture}}
\caption{Additional results of our method for diseased subjects, for $L=500$ and $s=100$.}
\end{figure}
\section{Additional Results on the Brats2020 Dataset}
\begin{figure}
\centering
\resizebox{0.875\textwidth}{!}{
	\begin{tikzpicture}
	\node[] at (0,3.85)  {\scriptsize T1};
	\node[] at (2.4,3.85)  {\scriptsize T1ce};
	\node[] at (4.8,3.85)  {\scriptsize T2};
	\node[] at (7.2,3.85)  {\scriptsize FLAIR};
	\node[] at (9.6,3.85)  {\scriptsize Anomaly Map};
	
	\node[draw=white, inner sep=0pt, thick] at (0, 2.4) {\includegraphics[scale=0.27]{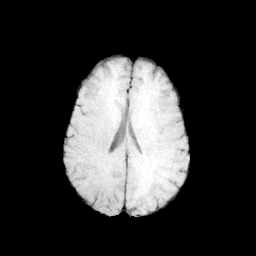}};
	\node[draw=white, inner sep=0pt, thick] at (2.4, 2.4) {\includegraphics[scale=0.27]{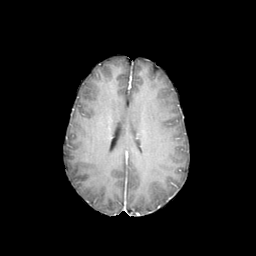}};
	\node[draw=white, inner sep=0pt, thick] at (4.8, 2.4) {\includegraphics[scale=0.27]{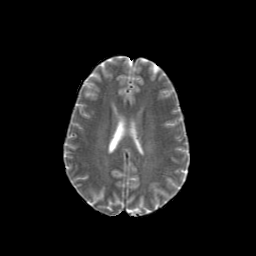}};
	\node[draw=white, inner sep=0pt, thick] at (7.2, 2.4) {\includegraphics[scale=0.27]{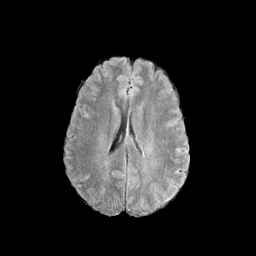}};
		\node[draw=white, inner sep=0pt, thick] at (9.6,2.4) {\includegraphics[scale=0.287]{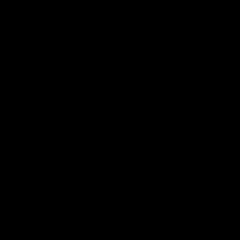}};
	\node[draw=white, inner sep=0pt, thick] at (0, 0) {\includegraphics[scale=0.27]{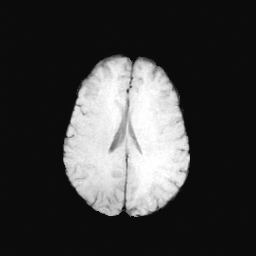}};
	\node[draw=white, inner sep=0pt, thick] at (2.4, 0) {\includegraphics[scale=0.27]{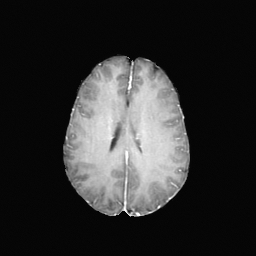}};
	\node[draw=white, inner sep=0pt, thick] at (4.8,  0) {\includegraphics[scale=0.27]{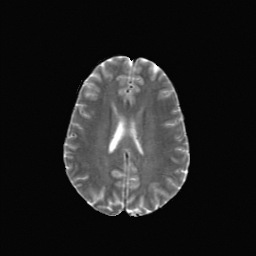}};
	\node[draw=white, inner sep=0pt, thick] at (7.2,  0) {\includegraphics[scale=0.27]{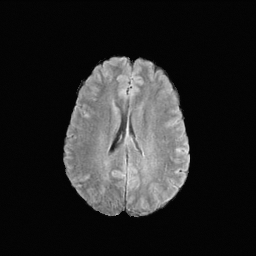}};
	\node[draw=white, inner sep=0pt, thick] at (9.6,0)  {\includegraphics[scale=0.187]{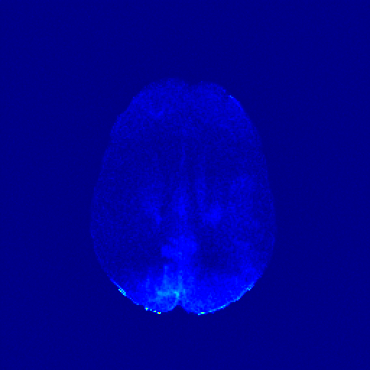}};
	\node[rotate=90]  at (-1.5,0) {\scriptsize Output};
	\node[rotate=90] at (-1.5,2.4) {\scriptsize Input};
	\end{tikzpicture}}
	\caption{Results of our method for a healthy subject, for $L=500$ and $s=100$.}
	
\end{figure}

\begin{figure}
\centering
\resizebox{0.92\textwidth}{!}{
	\begin{tikzpicture}
	\node[] at (0,18.8)  {\scriptsize T1};
	\node[] at (2.4,18.8)  {\scriptsize T1ce};
	\node[] at (4.8,18.8)  {\scriptsize T2};
	\node[] at (7.2,18.8)  {\scriptsize FLAIR};
	\node[] at (9.6,18.8)  {\scriptsize Anomaly Map};
	
	\node[draw=white, inner sep=0pt, thick] at (0, 2.4) {\includegraphics[scale=0.28]{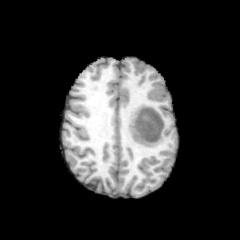}};
	\node[draw=white, inner sep=0pt, thick] at (2.4, 2.4) {\includegraphics[scale=0.28]{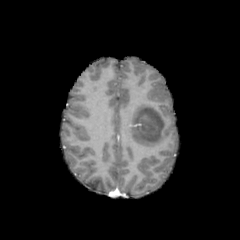}};
	\node[draw=white, inner sep=0pt, thick] at (4.8, 2.4) {\includegraphics[scale=0.28]{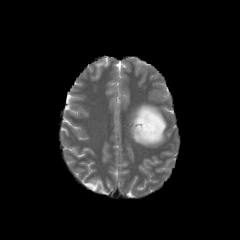}};
	\node[draw=white, inner sep=0pt, thick] at (7.2, 2.4) {\includegraphics[scale=0.28]{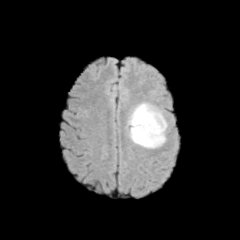}};
		\node[draw=white, inner sep=0pt, thick] at (9.6,2.4) {\includegraphics[scale=0.28]{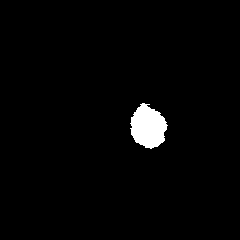}};
	\node[draw=white, inner sep=0pt, thick] at (0, 0) {\includegraphics[scale=0.28]{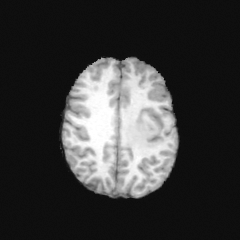}};
	\node[draw=white, inner sep=0pt, thick] at (2.4, 0) {\includegraphics[scale=0.28]{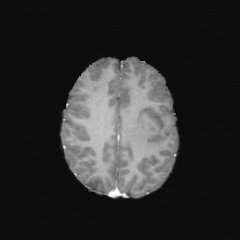}};
	\node[draw=white, inner sep=0pt, thick] at (4.8,  0) {\includegraphics[scale=0.28]{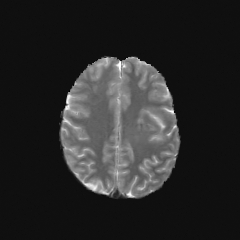}};
	\node[draw=white, inner sep=0pt, thick] at (7.2,  0) {\includegraphics[scale=0.28]{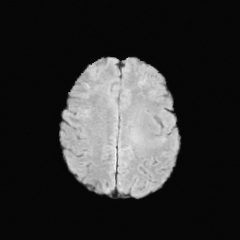}};
	\node[draw=white, inner sep=0pt, thick] at (9.6,0)  {\includegraphics[scale=0.182]{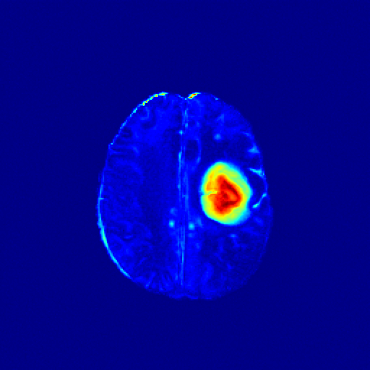}};
	\node[rotate=90]  at (-1.35,0) {\scriptsize Output};
	\node[rotate=90] at (-1.35,2.2) {\scriptsize Input};
	
		\node[draw=white, inner sep=0pt, thick] at (0, 7.4) {\includegraphics[scale=0.28]{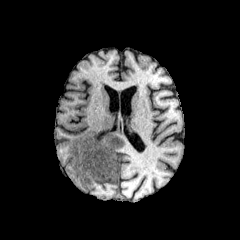}};
	\node[draw=white, inner sep=0pt, thick] at (2.4, 7.4) {\includegraphics[scale=0.28]{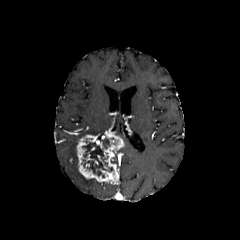}};
	\node[draw=white, inner sep=0pt, thick] at (4.8, 7.4) {\includegraphics[scale=0.28]{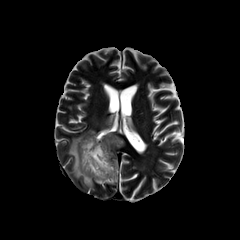}};
	\node[draw=white, inner sep=0pt, thick] at (7.2, 7.4) {\includegraphics[scale=0.28]{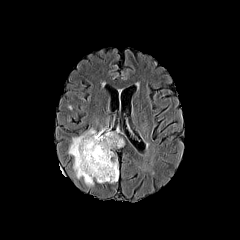}};
		\node[draw=white, inner sep=0pt, thick] at (9.6,7.4) {\includegraphics[scale=0.28]{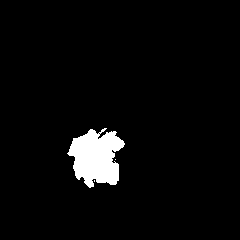}};
	\node[draw=white, inner sep=0pt, thick] at (0, 5) {\includegraphics[scale=0.28]{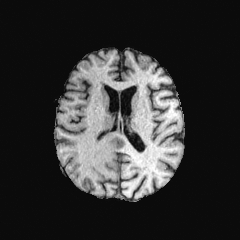}};
	\node[draw=white, inner sep=0pt, thick] at (2.4, 5) {\includegraphics[scale=0.28]{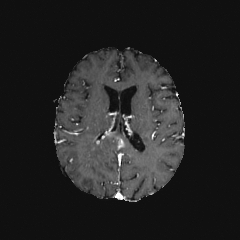}};
	\node[draw=white, inner sep=0pt, thick] at (4.8,  5) {\includegraphics[scale=0.28]{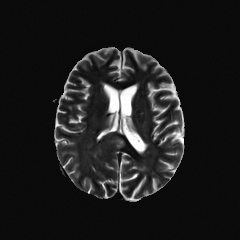}};
	\node[draw=white, inner sep=0pt, thick] at (7.2,  5) {\includegraphics[scale=0.28]{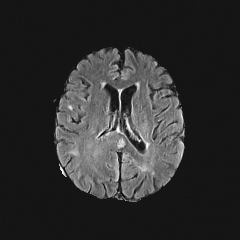}};
	\node[draw=white, inner sep=0pt, thick] at (9.6,5)  {\includegraphics[scale=0.182]{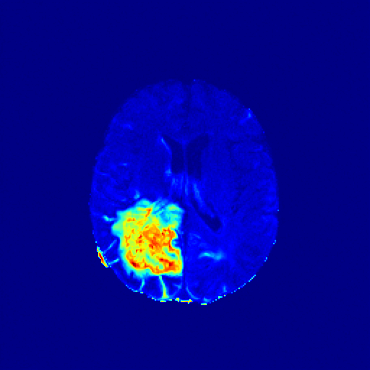}};
	\node[rotate=90]  at (-1.35,5) {\scriptsize Output};
	\node[rotate=90] at (-1.35,7.4) {\scriptsize Input};
	
		\node[draw=white, inner sep=0pt, thick] at (0, 12.4) {\includegraphics[scale=0.28]{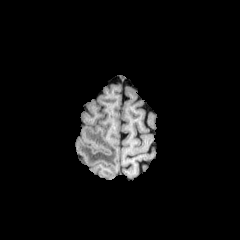}};
	\node[draw=white, inner sep=0pt, thick] at (2.4, 12.4) {\includegraphics[scale=0.28]{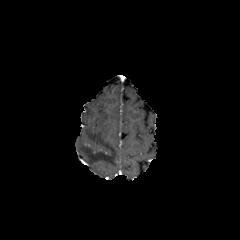}};
	\node[draw=white, inner sep=0pt, thick] at (4.8, 12.4) {\includegraphics[scale=0.28]{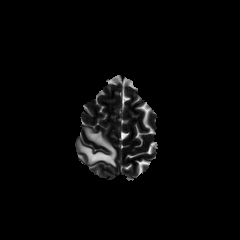}};
	\node[draw=white, inner sep=0pt, thick] at (7.2, 12.4) {\includegraphics[scale=0.28]{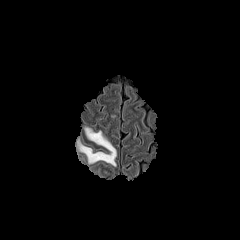}};
		\node[draw=white, inner sep=0pt, thick] at (9.6,12.4) {\includegraphics[scale=0.28]{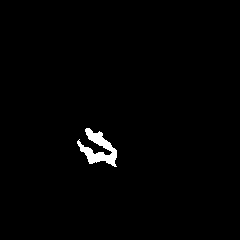}};
	\node[draw=white, inner sep=0pt, thick] at (0, 10) {\includegraphics[scale=0.28]{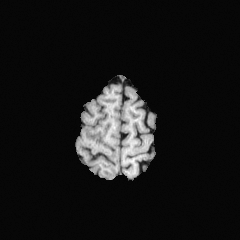}};
	\node[draw=white, inner sep=0pt, thick] at (2.4, 10) {\includegraphics[scale=0.28]{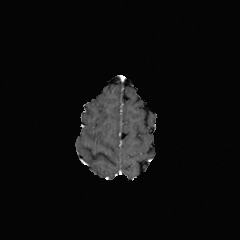}};
	\node[draw=white, inner sep=0pt, thick] at (4.8,  10) {\includegraphics[scale=0.28]{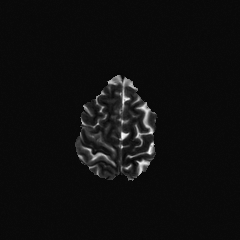}};
	\node[draw=white, inner sep=0pt, thick] at (7.2,  10) {\includegraphics[scale=0.28]{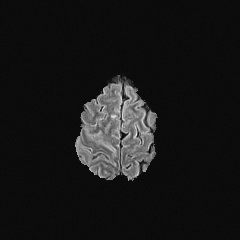}};
	\node[draw=white, inner sep=0pt, thick] at (9.6,10)  {\includegraphics[scale=0.182]{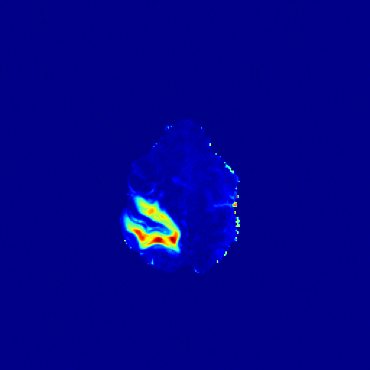}};
	\node[rotate=90]  at (-1.35,10) {\scriptsize Output};
	\node[rotate=90] at (-1.35,12.4) {\scriptsize Input};
	
			\node[draw=black, inner sep=0pt, thick] at (0, 17.4) {\includegraphics[scale=0.28]{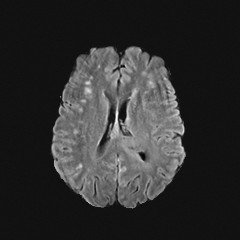}};
	\node[draw=white, inner sep=0pt, thick] at (2.4, 17.4) {\includegraphics[scale=0.28]{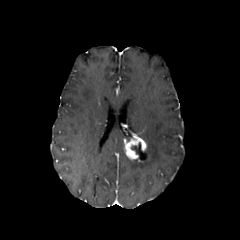}};
	\node[draw=white, inner sep=0pt, thick] at (4.8, 17.4) {\includegraphics[scale=0.28]{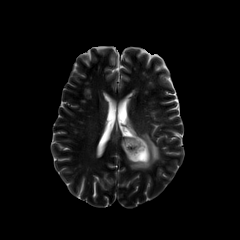}};
	\node[draw=white, inner sep=0pt, thick] at (7.2, 17.4) {\includegraphics[scale=0.28]{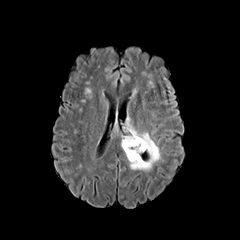}};
		\node[draw=white, inner sep=0pt, thick] at (9.6,17.4) {\includegraphics[scale=0.28]{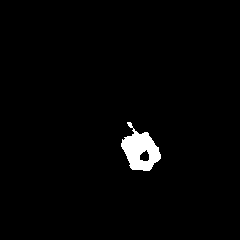}};
	\node[draw=white, inner sep=0pt, thick] at (0, 15) {\includegraphics[scale=0.28]{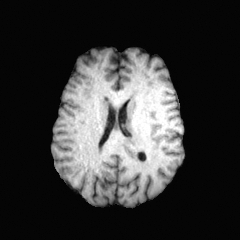}};
	\node[draw=white, inner sep=0pt, thick] at (2.4, 15) {\includegraphics[scale=0.28]{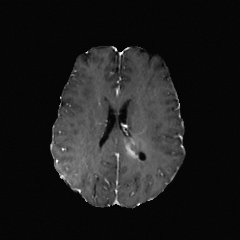}};
	\node[draw=white, inner sep=0pt, thick] at (4.8,  15) {\includegraphics[scale=0.28]{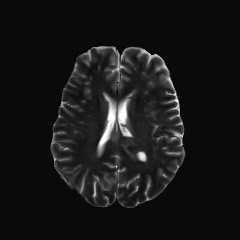}};
	\node[draw=white, inner sep=0pt, thick] at (7.2,  15) {\includegraphics[scale=0.28]{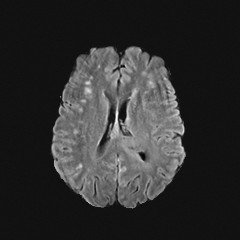}};
	\node[draw=white, inner sep=0pt, thick] at (9.6,15)  {\includegraphics[scale=0.182]{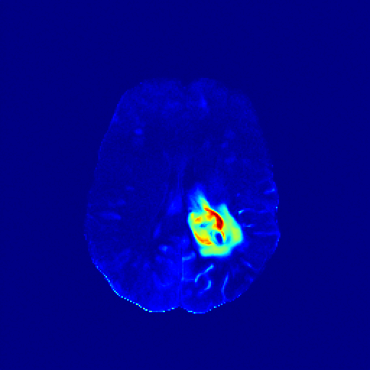}};
	\node[rotate=90]  at (-1.35,15) {\scriptsize Output};
	\node[rotate=90] at (-1.35,17.4) {\scriptsize Input};
	\end{tikzpicture}}
	\caption{Additional results of our method for diseased subjects, for $L=500$ and $s=100$.}
\end{figure}